\documentclass[aps,reprint,onecolumn,prfluids,11pt,tightenlines]{revtex4-2}
\usepackage{amsmath}
\usepackage{amssymb}
\usepackage{color}
\usepackage[utf8]{inputenc}
\usepackage[T1]{fontenc}
\usepackage{graphicx}
\usepackage{hyperref}
\usepackage{multirow}
\hypersetup{
    colorlinks=true,
    linkcolor=blue,
    filecolor=magenta,      
    urlcolor=cyan,
    citecolor=blue,
}
\usepackage{natbib}

\begin{document}

\title{Jet drop production from bubbles with neighbors}

\author{Tristan Aurégan$^1$, Noé Daniel$^{1,2}$, Megan Mazzatenta$^1$, Luc Deike$^{1,3}$}

\affiliation{$^1$ Department of Mechanical and Aerospace Engineering, Princeton University, Princeton, New Jersey, USA}
\affiliation{$^2$ Département de Physique, École Normale Supérieure, Paris, France}
\affiliation{$^3$ High Meadows Environmental Institute, Princeton University, Princeton, New Jersey, USA}

\date{\today}

\begin{abstract}
    Bubbles bursting at the surface of the ocean produce drops that heavily influence ocean-atmosphere interactions. One of the mechanisms through which drops are formed is called jet drop production, where the collapse of the bubble cavity leads to the formation of a fast upwards jet that breaks to form drops. While isolated bubble bursting has been extensively studied, bubbles are often found in rafts (for instance in the ocean surface or a sparkling wine glass) and the understanding of collective effects remains more limited. We investigate experimentally how jet drop formation is modified by the presence of neighboring bubbles during the collapse. With the help of multiple high speed views of the collapsing bubble, we show how a change of cavity shape during collapse leads to the selection of smaller, faster, and more numerous drops. The size of the emitted drops is monotonically reduced with increasing number of neighboring bubbles (up to six for hexagonal packing) with the size reduction reaching a factor 5. The drop size distribution associated with bubbles arranged in rafts of various sizes is therefore much wider than in the case of isolated bubbles, and with a peak shifted to smaller sizes.
\end{abstract}

\maketitle

\section{Introduction}

When a surface bubble at an air-liquid interface bursts, its cap quickly retracts, leaving an open cavity at the surface of the liquid. This cavity then collapses through the action of capillary forces, forming an upwards jet at its center, that then destabilizes into drops. These \emph{jet drops} are crucial to understand air-sea exchanges of heat \citep{veron_ocean_2015}, and for the generation of sea salt aerosols \citep{lewis_sea_2004,deike_mass_2022,deike_mechanistic_2022}. In addition, these drops are very efficient pathways for the aerosolization of chemical compounds \citep{ghabache_physics_2014,sampath_aerosolization_2019} or microplastics \citep{shaw_ocean_2023} that might be present in the subphase. 

For these reasons, jet drop production has been widely studied in experimental \citep{blanchard_size_1989,lee_size_2011,ghabache_physics_2014,ghabache_evaporation_2016,brasz_minimum_2018} and numerical works \citep{duchemin_jet_2002,deike_dynamics_2018,brasz_minimum_2018,lai_bubble_2018,berny_role_2020,berny_statistics_2021} leading to scaling theories to predict the size and velocity of the ejected drops \cite{ganan-calvo_revision_2017,ganan-calvo_scaling_2018,gordillo_capillary_2019,blanco-rodriguez_sea_2020}. For a bubble of radius $R_b$ with a liquid of density $\rho$, viscosity $\mu$, and surface tension $\gamma$, the governing non-dimensional numbers of the dynamics are the Bond ($\mathrm{Bo}$) and Laplace ($\mathrm{La}$) numbers:
\begin{equation}
    \mathrm{Bo} = \frac{R_b^2}{\ell_c^2} \quad{\rm and}\quad \mathrm{La} = \frac{\rho \gamma R_b}{\mu^2},
\end{equation}
with $g$ the gravitational constant, and $\ell_c = \sqrt{\gamma / \rho g}$ the capillary length. The Bond number compares the magnitude of gravitational and interfacial forces, while the Laplace number compares interfacial and viscous forces. At high Laplace numbers, the collapse of the cavity produces many capillary ripples that propagate downwards towards the center of the cavity and produce large and slow jets \citep{ghabache_physics_2014}. At lower Laplace numbers $\mathcal{O}(10^3)$ these capillary ripples are damped, leading to a self similar collapse of the cavity \citep{duchemin_jet_2002,ghabache_physics_2014,brasz_minimum_2018,lai_bubble_2018} and thin and fast jets. At even lower Laplace numbers, viscosity damps the ejection until a critical Laplace number $\mathrm{La}_\star \approx 500$ is reached, below which no drops are produced. These observations have allowed authors to find scalings that describe very well the first droplet size $R_{d_1}$ and velocity $V_{d_1}$ for Laplace numbers greater than $10^3$ \citep{ganan-calvo_scaling_2018,gordillo_capillary_2019,blanco-rodriguez_sea_2020,deike_dynamics_2018}. 

In an effort to bring the results above closer to practical applications, jet drop production by isolated bubbles has further been studied in the presence of contaminants. These contaminants can be oil layers at the interface \citep{yang_enhanced_2023a,kulkarni_bursting_2024,yang_jet_2025}, polymers \citep{rodriguez-diaz_bubble_2023,ji_secondary_2023,sanjay_bursting_2021,dixit_viscoelastic_2025} or surface-active agents \citep{pierre_influence_2022,constante-amores_dynamics_2021,neel_collective_2021,neel_role_2022,pico_surfactantladen_2024,vega_influence_2024,eshima_size_2025,rodriguez-aparicio_critical_2025}. These contaminants can drastically alter the size of the first ejected droplet, increasing or decreasing it depending on the regime of Laplace number considered, or even completely suppress drop production. 

Beyond directly altering the cavity collapse through interfacial properties, contaminants can also suppress the coalescence of surface bubbles. Bubbles therefore form rafts instead of coalescing into larger bubbles (this effect is particularly visible in the experiments of \citet{neel_role_2022,mazzatenta_linking_2025}). As a consequence, when present at the surface of contaminated water, or in a sparkling wine glass \citep{seon_effervescence_2017}, bubbles do not burst in isolation like in most of the previously cited studies. In reality, the collapse of the cavity occurs with other cavities present very close to the bubble. In this study, we therefore experimentally investigate the effect of neighboring bubbles on jet drop production. Assuming that any potential effect of the neighbors is largest when the bubble and its neighbors are about the same size, we consider only cases where all bubbles in a raft are the same size. This results in rafts of surface bubbles similar to experiments from \citet{neel_collective_2021} or \citet{neel_role_2022}, with bubbles naturally arranging in a hexagonal pattern at the surface. The present study focuses on millimetric bubbles in water, resulting in Laplace numbers of the order $10^5$. The bursting of bubbles with neighbors in hexagonal configurations has been studied numerically by \citet{singh_numerical_2019}, who hinted towards neighbors decreasing the size and increasing the velocity of the ejected drops, and experimentally by \citet{seon_effervescence_2017}, who observed a sucking of neighboring bubbles towards the collapsing cavity and a tilting of the jet towards the bubble-free direction in cases with fewer than six neighbors. 

Linking jet drop properties to the local bubble configuration around a collapsing cavity is experimentally challenging, as it requires simultaneously measuring information about the raft (number of neighbors, or raft size) and the jet drops (size, velocity, and number) with synchronized cameras. 

When studying bursting bubbles arranged in rafts, two distinct cases quickly emerge: either a single bubble in the raft bursts, or multiple bubbles burst in quick succession \citep{ritacco_lifetime_2007,neel_collective_2021}. These two different regimes will be studied in distinct sections of this paper: Sec. \ref{sec:single_burst} is dedicated to single bursting events, while Sec. \ref{sec:multi_burst} is dedicated to multiple bursting events. Whether two bursting events are considered correlated is defined according to their proximity in space and time, with specific thresholds detailed later on. 

The article is structured as follows: in Sec. \ref{sec:setup} we describe the different experimental setups used to measure drop size and raft decays, in Sec. \ref{sec:single_burst} we discuss the effects of neighboring bubbles on the number of drops and their size or velocity, in Sec. \ref{sec:multi_burst} we analyze the case of multiple bursting event and compute the expected drop size distribution over a raft monodisperse bubbles, and Sec. \ref{sec:conclusion} discusses how to use these results in the more general case from \citet{neel_role_2022}.

\section{Experimental methods}\label{sec:setup}

We discuss the experimental setup and the measurement techniques deployed to characterize the effect of neighbors on collapsing cavity and jet drop formation. 

\subsection{Setup}

\begin{figure}[htbp]
    \centering
    \includegraphics[width=0.9\linewidth]{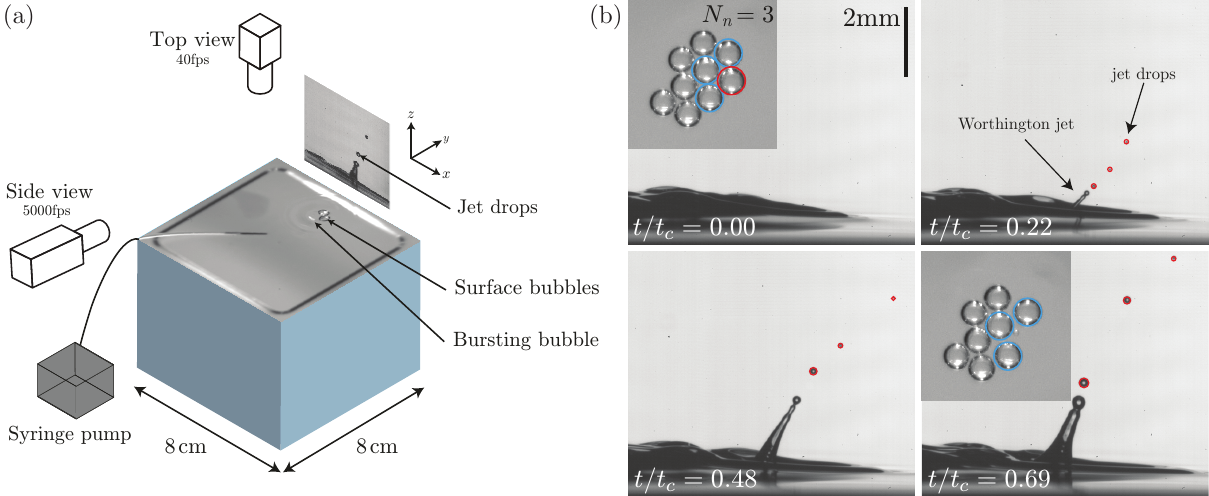}
    \caption{(a) Schematic of the experimental setup used to simultaneously measure surface bubbles and jet drops. A nearly monodisperse distribution of surface bubbles is generated by injecting air at the bottom of a water tank with a syringe pump. As air is released, bubbles rise to the surface where, under sufficient surfactant contamination, they persist and form bubble rafts. Surface bubbles are detected from top-view images and the jet drops detection relies on side-view high speed imaging. \quad (b) Example of the synchronized dual-view of a bursting event. The main view shows a tilted ejected jet destabilizing into drops (red). The times since the bursting of the cap are given in terms of the capillary timescale: $t_c=\sqrt{\rho R_b^3/\gamma} \approx 3.8$~ms. The inset in the top left panel shows the top view just before bursting with the bursting bubble highlighted in red and the neighbors in blue. The inset in the bottom right panel shows the raft on the next top view frame, with the bursting bubble having disappeared from the raft.}
    \label{fig:setup}
\end{figure}

We generate monodisperse rafts of bubbles in a cube tank with a side length of $8~$cm and let them decay until every single bubble has burst. The bubbles are generated by pushing air with a syringe pump through a small capillary, and we generate rafts of various sizes by changing the duration of the pulse of air.

To investigate bubble bursting dynamics and the formation of jet drops, we use a dual-camera imaging system that simultaneously captures two orthogonal views: a top view $(x,y)$ to monitor surface bubbles and detect bursting events, and a side view $(x,z)$ to resolve the subsequent jet and drop ejection (see Fig~\ref{fig:setup} (a)). We observe the bursting dynamics and jet drops ejection in the side view using a high speed camera (Phantom V2012, the pixel size in the most zoomed in case is 6~\textmu m) at 5000 frames per second. The top view is recorded at 40 frames per second (Basler acA1440-220um). Illumination for both imaging planes is provided by two independent LED light panels. Thanks to the short lifetime of the bubbles, several bursting events in a raft can be saved with a single recording of the high speed camera. Each of the bursting events is then manually selected on both cameras and separately saved.

We use mixtures of deionized water and Sodium Dodecyl Sulfate (SDS) for the solutions. This allows us to prevent coalescence, therefore enabling raft formation and allowing the bubbles to interact. We found that a minimum SDS concentration of $c=4~\mu\mathrm{mol}/L$ is necessary to prevent coalescence. In this study, we vary SDS concentration $c$ and bubble size $R_b$ (see Tab. \ref{tab:experiments}). All SDS concentrations used are very small and well below the Critical Micelle Concentration (or CMC, 8.2 mmol/L for SDS), resulting in a surface tension close to that of pure water ($\gamma=71\pm1$~mN/m). Bubble radii are found to follow a centered distribution around $\langle R_b \rangle$ with a standard deviation of $\pm 0.1\mathrm{mm}$. For each of these conditions, we recorded about 60 bursting events spanning many neighbor configurations in rafts of at most $\sim10$ bubbles. Using bubbles with a radius of around 1~mm in water means that the Laplace number is always of the order $10^5$ and the Bond number is about 0.3. In particular, this value of the Bond number results in bubbles that are mostly underwater with spherical cavities not deformed by the presence of neighbors \citep{toba_drop_1959,lhuissier_bursting_2012,yeom_correlations_2022}.

The above choice of contamination parameters means that jet drop production in the isolated case (i.e. with no neighbors) essentially behaves as predicted by the clean interface scalings from the literature \citep{deike_dynamics_2018,ganan-calvo_revision_2017,gordillo_capillary_2019}. The data for the isolated case (listed in Tab. \ref{tab:experiments}, last four columns) lies within previous experimental and numerical errors of drop Laplace number as a function of bubble Laplace number for a bubble without surfactants \citep{berny_role_2020}. \citet{pierre_influence_2022} describes a significant influence of SDS with a setup very similar to ours, but a large surfactant concentration ($\gtrsim 20\,\%$ of the CMC, to be compared to at most $0.3\,\%$ in our case) is required to see those changes.

\begin{table}[htbp]
    \centering
    \setlength{\tabcolsep}{7pt} 
    \renewcommand{\arraystretch}{1.1} 

    \begin{tabular}{ccccccccc}
    \hline
    $c$& 
    $c/\text{CMC}$&  
    $\langle R_b \rangle$& 
    $\# \text{bursting}$&
    \multirow{2}{*}{$\# \text{drops}$}&
    $R_{d_1}(N_n=0)$&
    $V_{d_1}(N_n=0)$&
    $\mathrm{La}$ &
    $R_{d_1} / \ell_\mu$ \\
    $(\mu\text{mol}/L)$ & 
    (\%) &  
    (mm) & 
    events&
    &
    ($\mu$m) &
    (m/s) &
    $(\times 10^4)$ &
    $(\times 10^4)$ \\
    \hline
    4.0   & 0.0488 & 1.08 & 68 & 181 & 246 & 2.19 & $10.3$ & $1.90$ \\
    13.6  & 0.166  & 0.88 & 61 & 240 & 266 & 2.06 & $12.2$ & $1.64$ \\
    13.6  & 0.166  & 1.23 & 81 & 183 & 225 & 2.06 & $11.0$ & $2.02$ \\
    13.6  & 0.166  & 1.29 & 82 & 184 & 234 & 2.16 & $11.4$ & $2.07$ \\
    20.4  & 0.249  & 1.39 & 63 & 273 & 231 & 2.12 & $10.9$ & $1.99$ \\
    20.4  & 0.249  & 1.26 & 58 & 206 & 188 & 2.52 & $7.80$ & $2.36$ \\
    27.2  & 0.332  & 1.18 & 62 & 195 & 219 & 2.16 & $9.90$ & $2.26$ \\
    \hline
    \end{tabular}
    \caption{Summary of experimental conditions and key measurements. $c$ is the concentration in SDS, $\left< R_b \right>$ is the average radius of the bubbles. Each experimental condition is repeated several times, and we report the number of single bursting events as well as the number of measured drops. $R_{d_1}(N_n=0)$ is the typical radius of the first ejected drop in the case with no neighbors, and $V_{d_1}(N_n=0)$ is the associated velocity. The jetting process is governed by the bubble Laplace number $\mathrm{La}=\rho\gamma R_b/\mu^2$ (order $10^5$) and results in a dimensionless drop size (in the isolated case) $R_{d_1}(N_n=0) / \ell_\mu$ with $\ell_\mu = \mu^2 / (\rho \gamma)$, of the order $10^4$.
    }
    \label{tab:experiments}
\end{table}

\subsection{Detection of bubbles and drops}

For each bursting event, the processing of the data consists of locating the bubble that bursts on the top view and measuring its radius $R_b$, its number of neighbors $N_n$, and the size of the raft it belongs to $\mathcal{R}$. In addition, for the same event we use the side view to measure jetting properties such as the radii $R_d$, velocities $V_d$ and number of the ejected drops $N_d$. This will lead to data sets on the size, velocity, and number of jet drops as a function of the number of neighbors. Note that in the case of multiple bursting events, these operations are repeated for each of the bubbles, with each of the jets associated with its corresponding bubble, and the bursting order in the sequence is also measured. 

We analyze the top view with a Convolutional Neural Network (CNN) based on the YOLO ("You Only Look Once") architecture \citep{redmon_you_2016}, using the open-source \texttt{ultralytics} \citep{glenn_ultralytics_2024} implementation with pretrained weights as a starting point. The CNN is fine-tuned on a custom dataset of 300 manually annotated top-view images. The training is performed on a \texttt{NVIDIA A100 40GB} GPU for about an hour, while inference on one picture using the same GPU takes less than $10\mathrm{ms}$. Results are returned as bounding boxes that tightly enclose each detected bubble. Given the nearly circular shape of the surface bubbles, we approximate each bubble's radius and center using the inscribed circle within the corresponding bounding box. This method provides accurate geometric information while retaining computational efficiency.

The performance of the trained model is highly satisfactory and similar to that of traditional techniques such as Canny edge detection and the Circular Hough Transform. The advantage of this approach is the superior robustness to noise and background variability, and the relatively fast inference across image sequences, in contrast to the higher computational cost associated with classical detection pipelines.

The positions and velocities of each of the detected bubbles are then used to track each individual bubble over time for the few frames before and after the bursting event, allowing us to detect which bubble has burst. From the final frame before rupture, we extract the bubble's radius and position, determine the number of adjacent neighbors $N_n$, and evaluate the raft size $\mathcal{R}$ of the cluster to which it belongs. In parallel, the radii of all detected bubbles are stored for each experiment, enabling us to construct the probability density function (PDF) of the bubble radius $R_b$ for every experimental condition. 

Two bubbles $i$ and $j$ are considered neighbors if the distance between their detected centers is below some critical value $\delta_{\rm max} R_b$. $\delta_{\rm max}$ is a dimensionless parameter chosen \textit{ad hoc}: in a monodisperse raft with hexagonal packing of bubbles, two neighboring bubbles give a dimensionless distance of $2$, while the next closest bubble sets the maximum distance $2\sqrt3 \approx 3.464$. For the whole study, we use $\delta_{\rm max}=2.3$. The raft corresponds to the largest connected ensemble of bubbles containing the currently bursting bubble. 

The side view high speed images are analyzed using conventional image processing techniques \citep{neel_velocity_2022}: thresholding and segmentation allows us to detect the drops on each frame. The drops are then tracked across frames, allowing us to measure each drop's size and velocity (see Fig. \ref{fig:setup} (b)). 

We imaged the collapse of the bubble cavity for qualitative analysis. This is done by using the high speed camera with a high framerate (20\,000 fps), and pointing it at the bubble from underneath the surface or from the top of the bubble instead of the side. These videos are used in Sec. \ref{sec:cavities}.

\subsection{Statistical analysis of raft decay}\label{sec:setup_stat}

We also measured statistical properties of raft decay to obtain the average number of neighbor of bursting bubbles, or relative fraction of single versus multiple bursting events for instance. In order to obtain robust statistics, we focus in this case on few conditions ($c=13.6$ and 27.2 \textmu mol/L, and $\left< R_b \right>=1.3$~mm), and only recorded raft decay with the top view. This allows a large number of rafts ($\sim 100$) to be processed and analyzed from the generation to the last bubble bursting. The top view setup and processing is identical to the previous case. The mean radius in all the conditions tested is $\left< R_b \right> = 1.2$~mm, and we used three SDS concentrations: $13.6$, $27.2$, and $40.8~\mu\mathrm{mol}/\mathrm{L}$. The results of this analysis are discussed in Sec. \ref{sec:stat_decay}.

\section{Jet drop production in the presence of neighbors}\label{sec:single_burst}

This section is dedicated to the analysis of the modifications that occur to the jetting process when a single bubble bursts, in the presence of neighbors.

\subsection{Qualitative changes to jet drop production}

\begin{figure}[htbp]
    \centering
    \includegraphics[width=\linewidth]{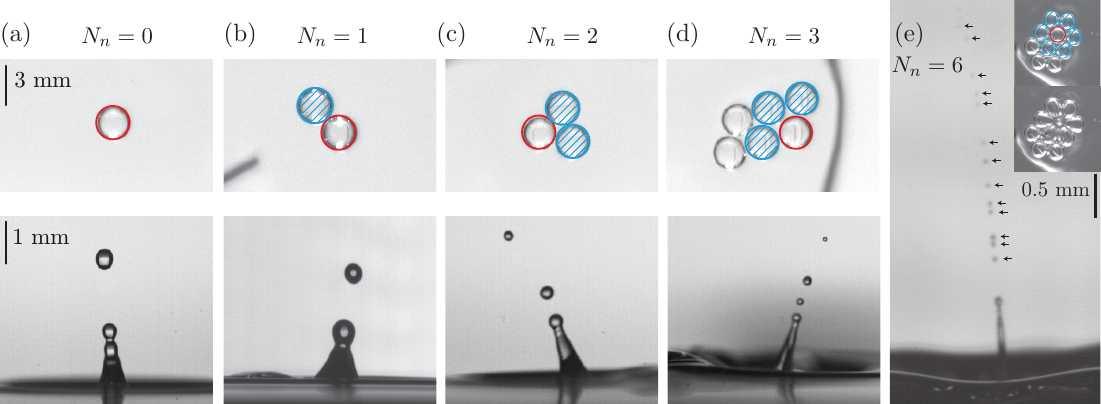}
    \caption{Variations of the jetting process as the number of neighbors is increased. Each panel (a) to (d) shows the top view just before the bubble bursts (top row, bursting bubble in red and neighbors in blue), and the side view as the first drop is ejected (bottom row). As the number of neighbors is increased, the jet can be seen getting thinner and tilting towards the side free of bubbles. (e) Example of bubble bursting with six neighbors and producing a very large number of droplets. Note that the scale is smaller in this case. The inset shows the top view just before (top) and after (bottom) bursting.}
    \label{fig:qualitative_jet}
\end{figure}

We observe that the presence of neighbors significantly modifies the drop size: when the number of neighbors $N_n$ increases, the size of the first drop $R_{d_1}$ (and the width of the overall jet) decreases. This is illustrated in Fig. \ref{fig:qualitative_jet} with examples of jet drop formed for $N_n$ going from 0 (isolated bubble) to 6 (hexagonal packing). At the same time, the jet tilts towards the direction free of neighboring bubbles. Both of these observations are coherent with previous qualitative observations of the same problem \citep{seon_effervescence_2017,singh_numerical_2019}. The tilting of the jet has been linked with the breaking of the symmetry of the initial configurations when $0 < N_n < 6$. The mechanism for the drop size reduction has never been explained and is discussed in Sec. \ref{sec:cavities}.

When enough monodisperse bubbles reach the surface, they arrange in a hexagonal packing so that $6$ is the maximum value of $N_n$. The reduction of the jet drop size is then striking. 
One example of this configuration is shown in Fig. \ref{fig:qualitative_jet} (e), with corresponding top views before and after the bursting. These images highlight a remarkably high drop production, in contrast to typical cases with smaller numbers of neighbors shown in Fig. \ref{fig:qualitative_jet} (a) to (d). The resulting jet appears extremely thin, and the drops ejected are both numerous and very small compared to the one formed by an isolated bubble. The top-view sequences also exhibit a characteristic pattern previously documented by \citet{seon_effervescence_2017}. Using high-speed imaging, they demonstrated that in hexagonally packed bubble rafts, the bursting of a central bubble induces a radial inward motion of the surrounding bubbles. This motion forms a transient flower pattern, resulting from the suction effect created by the rapid collapse of the bursting cavity.

Some symmetry is recovered in the case with six neighbors and as a consequence the jet is not tilted anymore, while the drop size reduction persists. Note that there is high variability both in the size of the jet and the number of droplets ejected within observations of bubbles bursting with six neighbors: the case presented in Fig. \ref{fig:qualitative_jet} (e) is striking in the number of drops ejected, and the droplet size is at the limit of the resolution of our camera, but other cases have slightly fewer, larger drops (see below).

\subsection{Jet drops properties with increasing neighbors}\label{sec:quant_drops}

\begin{figure}[htbp]
    \centering
    \includegraphics[width=0.6\linewidth]{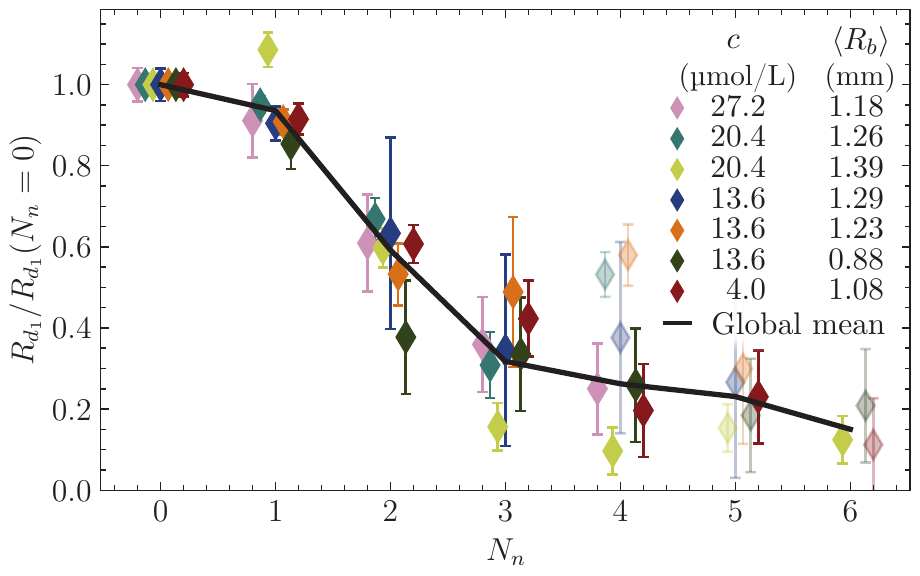}
    \caption{Evolution of the normalized radius of the first drop, $R_{d_1}/R_{d_1}(N_n=0)$ as a function of the number of adjacent neighbors $N_n$, for various contamination levels $c$ and surface bubble sizes $\left< R_b \right>$. Each datapoint corresponds to several repetitions with a given number of neighbors. The marker shows the mean, and the errorbar length is one standard deviation, with a minimal value corresponding to 4 pixels. Data points shown with transparency indicate cases where fewer than four measurements were available to compute the mean. The black line corresponds to the average of all cases.}
    \label{fig:first_drop_rad}
\end{figure}

We now quantitatively analyze the evolution of the size of the first ejected drop as the number of neighbors increases (see Fig. \ref{fig:first_drop_rad}). We see a gradual decrease of the size for all cases (SDS concentration and bubble radius) tested. The size of the droplet (and of the jet) decreases very rapidly between cases with 1 to 3 neighbors. For cases with 4 to 6 neighbors, the size of the droplets can be up to five times smaller than in the isolated bubble case. This large variation of size is of the same order of magnitude as the maximum effect of surfactants \citep{pierre_influence_2022,eshima_size_2025} or oil coatings \citep{yang_enhanced_2023a} at a fixed Laplace number. The size of the first droplet in the case with no neighbors ($R_{d_1}(N_n=0)$) varies by about 20\% (from changes in SDS concentration or bubble radius, see Tab. \ref{tab:experiments}) yet the trend with $N_n$ is remarkably similar across cases, suggesting that the size decrease that we observe is not strongly dependent on the parameters of the experiment within the range that we study. 

To give actual numbers, with bubbles of about 1.2~mm in radius, the average drop size across all cases for $N_n=0$ is 230~microns (in agreement with the expected scalings), and the drop size when $N_n=6$ is reduced to about 30 microns. 

\begin{figure}[htbp]
    \centering
    \includegraphics[width=0.8\linewidth]{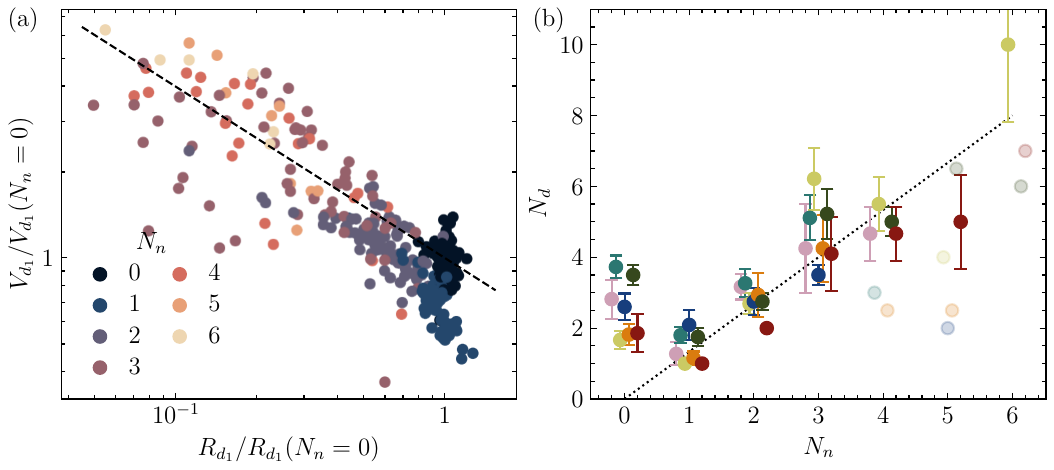}
    \caption{(a) Relation between the first drop's radius and velocity, both normalized by the case with no neighbors. The data shown regroups all cases of the study, and the color indicates the number of neighbors. The dashed line represents combined scalings from the literature (i.e. $V_{d_1} \propto R_{d_1}^{-3/5}$), which is consistent with the present data (Eq.~\ref{eq:velrad}). (b) Evolution of the number of drops ejected with the number of neighbors. Colors represent different cases identically to Fig. \ref{fig:first_drop_rad}. Markers show the mean, and the errorbar length is one standard deviation. Data points shown with transparency indicate cases where fewer than four measurements were available to compute the mean. The dotted line is the trend line $N_d = 4 / 3 N_n$.}
    \label{fig:velrad}
\end{figure}

Another descriptor of the jetting dynamics is the velocity of the first drop. In the isolated case, the first drop radius and velocity can be linked by combining scalings for the radius and velocity from the literature \citep{ganan-calvo_revision_2017,deike_dynamics_2018,berny_role_2020}:
\begin{subequations}\label{eq:velrad}
    \begin{gather}
            \frac{R_{d_1}}{\ell_\mu} = k_r \varphi^{5/4} \quad {\rm and} \quad  \frac{V_{d_1}}{V_\mu} = k_v(\mathrm{Bo}) \varphi^{-3/4},\\
            \varphi = \sqrt{\mathrm{La}}\left(\sqrt{\frac{\mathrm{La}}{\mathrm{La}_\star}} - 1 \right),
    \end{gather}
\end{subequations}
where $\ell_\mu = \mu^2 / (\rho \gamma)$ and $V_\mu = \gamma / \mu$ are reference length and velocity scales, respectively. $k_r$ and $k_v = 19(1 + 2.2 \mathrm{Bo})$ are two dimensionless constants (fitted on isolated bubble bursting data in the literature), and $\mathrm{La}_\star\approx 500$ is the critical Laplace number below which no drops are produced. 

The relation between first drop size and velocity is plotted in Fig. \ref{fig:velrad} (a). As the number of neighbors is increased (colors), not only is the radius reduced, but the velocity of the jet is also increased. Despite the large dispersion of the data, the general trend of the relationship between velocity and radius is reasonably well captured by combining the scalings in Eq. \eqref{eq:velrad}, to show the evolution of the velocity as the radius changes (black dashed line). One outlier to this trend is the jump between 0 and 1 neighbors where the data indicates a slight decrease in size, but also a reduction in velocity, which is opposite to what the scaling above predicts. Note that $V_{d_1}$ is the norm of the drop velocity (not the vertical velocity) such that a tilting of the jet does not result in a reduction of velocity. Overall, increasing the number of neighbors therefore results in faster and thinner jets, in a manner which is similar to reducing the bubble Laplace number.

Increasing the number of neighbors also has a strong effect on the total number of drops produced (Fig. \ref{fig:velrad} (b)). Going from 1 to 6 neighbors increases the average number of drops from 2 to around 8 to 10, following a trend roughly captured by the equation $N_d = 4/3 N_n$ (dotted line). Note however that the number of droplets produced has a larger intrinsic uncertainty than the velocity and radius of the first drop and can strongly vary for a given condition. Interestingly, there is a jump between 0 and 1 neighbor, where the number of drops is actually slightly reduced when a neighbor is present. This specificity of the 1 neighbor case compared to the isolated one that was also visible on the velocity - size plot may be due to the breaking of the symmetry of the problem. This observation of an increased number of drops is also consistent with the idea that increasing the number of neighbors has a similar effect to reducing the Laplace number as the jets produced by low Laplace number bubbles also typically produce a large number of drops \citep{ghabache_surface_2015,berny_role_2020}.

After these initial observations, the effect of neighbors on the jetting properties seems to be similar to reducing the bubble Laplace number, resulting in smaller, faster, and more numerous droplets being ejected. The effect of neighbors on bubble bursting is surprisingly important, with a factor up to 5 of size reduction, around 4 of velocity increase and up to 5 times the number of drops. As a consequence, a 1 mm bubble in water ($\mathrm{La}\sim 10^5$) fully inside of a raft produces drops that are similar to what would be produced by an isolated 200 \textmu m bubble in water ($\mathrm{La}\sim 2\times 10^4$). To investigate the mechanism behind the modification of the jetting process, we focus in the following on high-speed recordings of cavity collapse. 

\subsection{Mechanisms for size reduction}\label{sec:cavities}

The production of jet drops is governed by the dynamics of cavity collapse following film rupture. In the isolated-bubble case, this process, whether in clean or contaminated conditions, has been investigated both numerically and experimentally. In this section, we aim to qualitatively compare these results to our experimental observations involving neighboring bubbles.

\begin{figure}[htbp]
    \centering
    \includegraphics[width=0.95\linewidth]{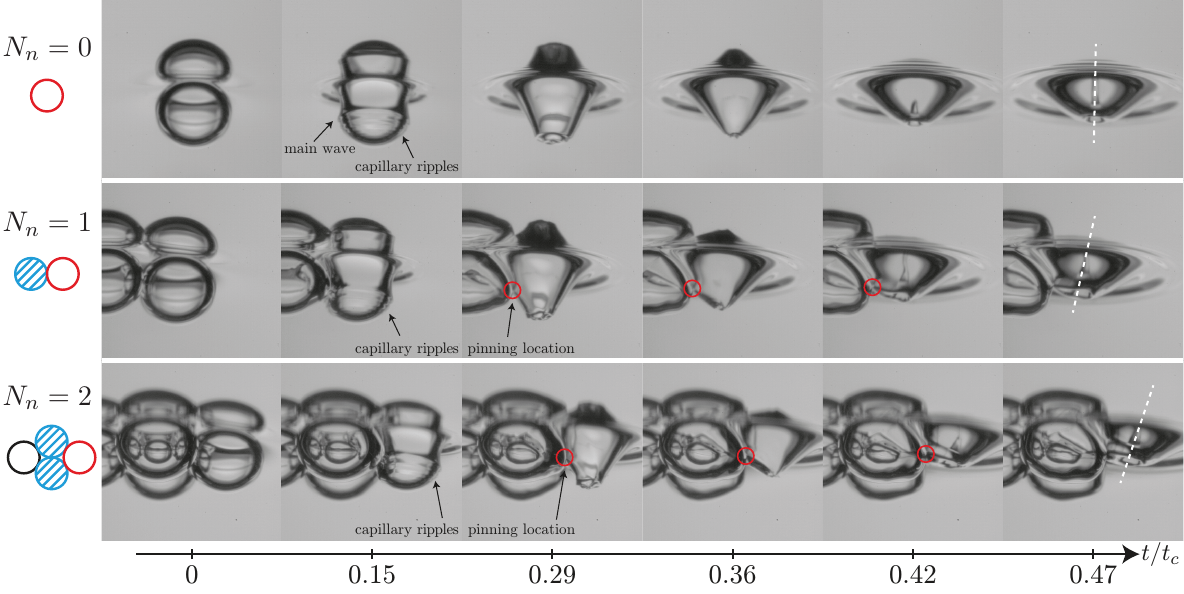}
    \caption{Chronophotography of side-view images showing cavity collapse dynamics for different neighboring configurations, all recorded at an SDS concentration of $c = 27.2~\mu\mathrm{mol}/\mathrm{L}$. A schematic of the top view is displayed for each case, with the red bubble bursting. Time of acquisitions are normalized by the capillary time $t_c = 3.8$~ms. In the presence of neighboring bubbles, the axisymmetry of the cavity collapse is broken, and the cavity profiles differs from the isolated case in particular on the side with neighbors.}
    \label{fig:bottom_cavities}
\end{figure}

When an isolated bubble bursts, capillary waves travel along the cavity wall towards the bottom of the bubble (Fig. \ref{fig:bottom_cavities}, top row). These waves can be divided into two groups: the large amplitude main capillary wave that leads to jet formation, and ripples travelling ahead of this main wave \citep{ghabache_physics_2014}. At low bubble Laplace numbers $\mathcal{O}(10^3)$, the ripples are attenuated by viscosity before they reach the bottom of the cavity and the collapse of the main wave is self-similar just before and after the main wave reaches the bottom of the cavity \citep{lai_bubble_2018}, leading to thin and fast jets. At higher Laplace numbers $\mathcal{O}(10^5)$, like in the present case, the capillary ripples reach the bottom of the cavity before the main wave, breaking self-similarity and leading to slower and larger jets. 

The presence of neighbors strongly disrupts the collapse of the cavity (Fig.~\ref{fig:bottom_cavities}, middle and bottom rows). The axisymmetry of the collapse is broken: on the side without neighbors, the capillary waves can be seen propagating almost as in the isolated case, while on the side facing neighboring bubbles important deviations from this base state occur. 
In the case with one neighbor only (middle row), the point on the cavity that is closest to the neighboring bubble (marked with a red circle) can be seen remarkably stable in space throughout the collapse, as if pinned by the neighboring interface. This pinning results in a tilting of the cavity bottom and horizontal momentum being imparted to the jet (to the right in this case). A similar phenomenon can be seen with two neighbors (bottom row), but this time the pinning location is located between the two neighboring bubbles. Once again the jet can be seen tilting away from the neighboring bubbles. 

\begin{figure}[htbp]
    \centering
    \includegraphics[width=0.99\linewidth]{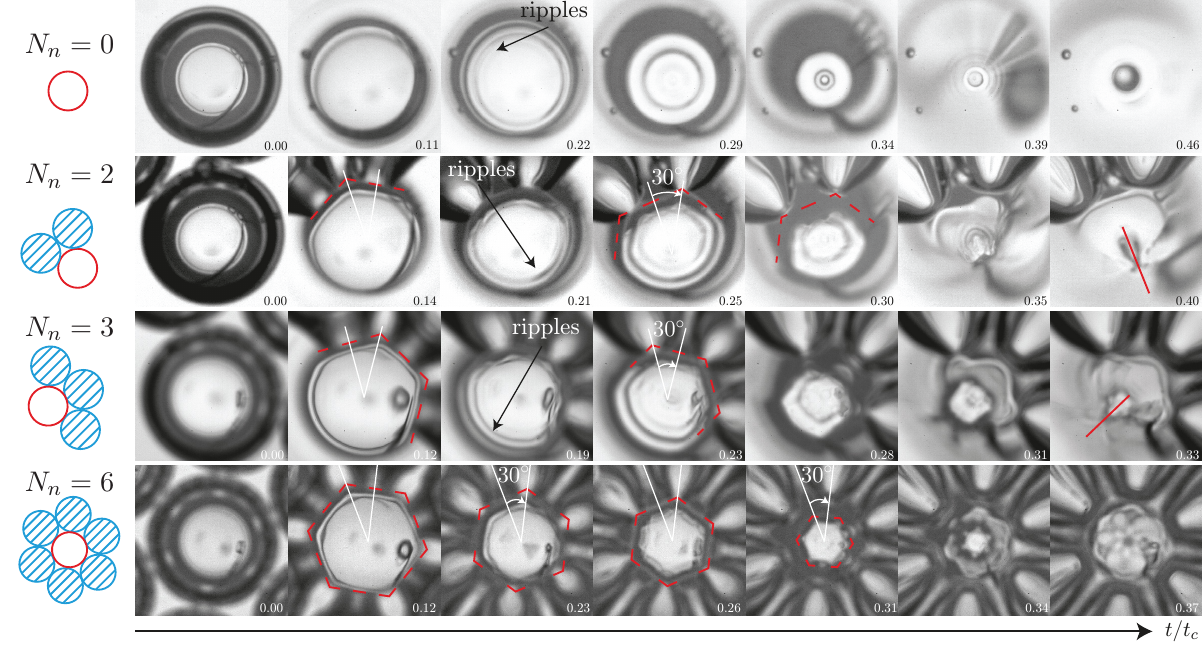}
    \caption{Chronophotography of top-view images showing cavity collapse under various neighboring configurations (similar conditions than in figure 5, $c = 27.2~\mu\mathrm{mol}/\mathrm{L}$). A schematic of the top view is displayed for each case, with the red bubble bursting. Time of acquisitions are normalized by the capillary time $t_c = 3.8$~ms. In the isolated case, the cavity collapse exhibits axisymmetric behavior, with surface waves propagating on a circular geometry. In contrast, when neighboring bubbles are present, the cavity adopts a polygonal shape, typically hexagonal, over part or all of its structure. }
    \label{fig:topdown_cavities}
\end{figure}

The presence of neighboring bubbles prevents the observation of the collapse from the side for higher neighbor counts. As a consequence, we turn to observations from the top of the cavity (Fig.~\ref{fig:topdown_cavities}) to see how the patterns of cavity collapse change. In the isolated case (top row), the collapse of the bubble is axisymmetric and capillary ripples are made visible by the circular, concentric shaded patterns they create. The main wave corresponds to a dark circle collapsing towards the center because of the very high curvature of this region. The presence of neighbors creates striking features in the cavity collapse: sharp corners appear in an otherwise spherical shape. At the beginning of the collapse ($t/t_c \approx 0.12$, bottom three rows), sharp 60 degree angles appear in the directions between neighboring bubbles. The segments between those angles are completely straight while the rest of the cavity in the directions with no bubbles remains spherical. For hexagonal packing $N_n=6$ (bottom row), this means that the cavity as seen from the top takes the shape of a hexagon. 

At later times, the sharp angles appear again with different orientations: for $N_n=2$ (Fig.~\ref{fig:topdown_cavities} second row) for instance, the fourth panel shows that the angles are in the directions towards the center of neighboring bubbles instead of towards the gap between bubbles. In the case with $N_n=6$, several rotations by $\pi / 6$ of the hexagonal pattern can be seen from the top before the waves reach the bottom of the cavity. The formation of polygonal shapes from N-fold symmetry, as well as successive rotations by $\pi / N$ also appear in the context of imploding shock waves \citep{eggers_singularities_2015a}. In this context, rotations are a consequence of the geometry of the problem and the velocity of the shocks. These similarities suggest a possible analogy between those two problems.

We can also see that the jet is always directed away from the bubbles in the cases with $0 < N_n < 6$.
In addition to strongly deforming the shape of the cavity during collapse, the presence of neighboring bubbles seems to also disrupt the propagation of the capillary ripples. This is particularly visible in the cases with $N_n=2$ and 3 (Fig.~\ref{fig:topdown_cavities} middle two rows) on the fourth panel, where ripples are clearly visible on the circular, neighbor-free side, but seem to be completely absent on the sides with neighbors from this view. 

\subsection{Jet selection in the presence of neighbors}

In summary, the presence of neighbors gradually increases the number of ejected drops, reduces their size and increases their velocity as the number of neighbors is increased. The variation in the jet shape is similar to what would occur for lower Laplace numbers, where viscosity damps the capillary ripples ahead of the main wave. Observations of the cavity show a strongly disrupted shape compared to the isolated case, with the bubble pinned to its neighbors in several locations, and the cavity forming 60 degree angles instead of a round shape. 

Overall, these observations are consistent with neighboring bubbles adding dissipation during the collapse of the cavity and the propagation of the waves. Each added neighbor increases the dissipation, resulting in the selection of thinner and faster jets as in the case of an isolated bubble. The cause for this added dissipation may be the deformation of neighboring bubbles during the collapse (the neighboring bubbles are sucked inwards as the cavity collapses, forming a \emph{flower pattern}, this effect has been reported by \citep{ritacco_lifetime_2007,seon_effervescence_2017} and is visible in Fig.~\ref{fig:topdown_cavities} bottom-right panel) or increased shear in the interstitial liquid between the bubbles. 

All trends observed so far are consistent with this interpretation, except for the case $N_n=1$. The velocity of the first drop is reduced instead of increased, and the number of ejected drops is slightly reduced (Fig.~\ref{fig:velrad}). These deviations may be interpreted as the need for two adjacent bubbles (and of the gap between them) to obtain the effect described above.

\section{Multiple bursting and raft evolution}\label{sec:multi_burst}

This section is dedicated to investigating the case where multiple bubbles in the same raft burst simultaneously, in contrast to the previous case where only one bubble was ever bursting at a time. Our goal is to see if the trends with respect to size and velocity of the drops previously described hold in this case, and to describe the expected drop production for a whole raft of bubbles. 

\begin{figure}[htbp]
    \centering
    \includegraphics[width=0.6\linewidth]{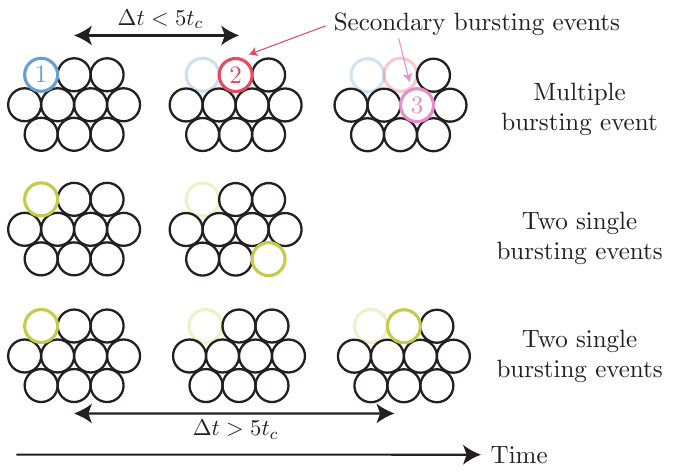}
    \caption{Types of events that are considered correlated. A bursting event is labeled as a multiple bursting event only if bursting bubbles are close in space and in time. Only direct neighbors are considered, and two events are considered correlated if they are separated in time by less than five capillary times. Bubbles that burst in a sequence, and that are not the first one are labeled as secondary.}
    \label{fig:schematic_multi}
\end{figure}

We consider that two bubbles bursting form a single bursting event if they are close both in space and in time (see Fig.~\ref{fig:schematic_multi}): the limit in space is that the two bubbles must be directly adjacent (i.e. the distance between their centers is less than $\delta_{\rm max} R_b$ with $\delta_{\rm max} = 2.3$, as defined in Sec.~\ref{sec:setup}), and the limit in time is five capillary timescales $t_c$ between the beginning of the two cap ruptures. This means that two bubbles bursting simultaneously on opposite sides of a large raft are considered not to influence each other during jet formation (see second example in Fig.~\ref{fig:schematic_multi}), but that a chain of correlated adjacent bursting events can form across long distances. The case where a single bubble bursts in a raft is designated as a `single bursting event' and the case where multiple bubbles burst in a correlated manner is called a `multiple bursting event'.

\subsection{Jet drop properties during multiple bursting events}

\begin{figure}[htbp]
    \centering
    \includegraphics[width=0.9\linewidth]{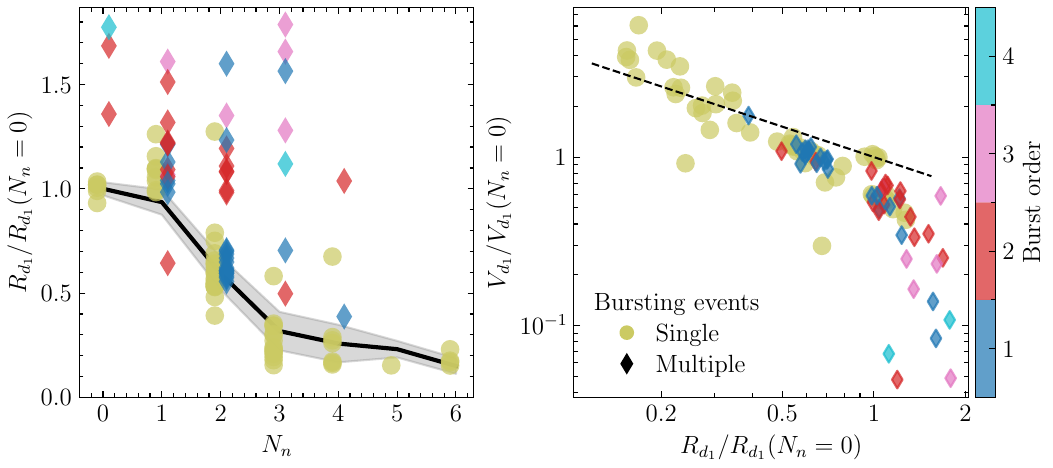}
    \caption{Effect of multiple bursting events on first drop radius and velocity compared to the trends established in the case of single bursting events. (a) Evolution of the normalized first drop radius $R_{d_1}/R_{d_1}(N_n=0)$ with the number of neighbors. Data with round markers corresponds to single bursting events, while diamonds correspond to multiple bursting events in a sequence, in which case the color corresponds to the order of bursting. Each marker is one bubble bursting event. The black line and shaded area correspond to the average and standard deviation of the single bursting event data in Fig.~\ref{fig:first_drop_rad}. (b) Velocity / size relationship with the same data and formatting. The dashed line corresponds $V_{d_1} \propto R_{d_1}^{-3/5}$ (Eq.~\eqref{eq:velrad}). All presented data is measured with $c=20.4~\mu\mathrm{mol}/L$ and $\langle R_b \rangle = 1.4~\mathrm{mm}$.}
    \label{fig:data_multi}
\end{figure}

We begin by analyzing the effect of multiple bursting events on the trends of velocity and size established in the previous section. For a given bubble size ($\langle R_b \rangle = 1.4~\mathrm{mm}$) and surfactant concentration ($c=20.4~\mu\mathrm{mol}/L$), we report both the data for single bursting events and multiple bursting events (Fig.~\ref{fig:data_multi}). The normalized radius as a function of number of neighbors is shown in panel (a): the black line represents the global average across all cases (surfactant concentrations and bubble sizes), and it is closely followed by the single bursting event data (round dark yellow markers, same data as in Fig.~\ref{fig:first_drop_rad} above). The data with multiple bursting event (diamonds) is color-coded by the burst order in the sequence (see Fig.~\ref{fig:schematic_multi}):  the bubbles bursting first in the sequence (dark blue) remain close to the single bursting event trend while the subsequent events show a significant spread. The subsequent events do not show any particular trend and produce drop radii that can either be increased or decreased by 50\% from the isolated case.

Similarly, in (b) we show the relationship between velocity and size of the first ejected drop. The first bursting event in a multiple bursting sequence (dark blue) lies within the single bursting event data (dark yellow). In contrast, the following events deviate from the trend and can reach very slow velocities. 

Overall, in multiple bursting events, the first bursting bubble seems to be relatively unaffected by other bursting bubbles. This result makes sense as, unless two bubbles burst simultaneously, the waves propagating in the cavity of the first one have time to form a jet before being affected by the collapse of the neighboring bursting bubble. From high speed observations, it seems to be the waves propagating outwards from the first bubble that trigger secondary bursting events, as noted by \citep{ritacco_lifetime_2007}. The following bursting events are heavily perturbed by the first one: cavities are deformed before they start collapsing, resulting in the loss of all coherence with the bursting order or the number of neighbors. 

These results then prompt the question: how many of the bubbles in a raft typically burst in multiple bursting events compared to single bursting events? In other words how much is the large size reduction described in Sec.~\ref{sec:single_burst} visible in the overall drop size distribution generated by a whole raft? To answer these questions, we analyze statistically the decay of rafts in the following section, and construct an estimator of raft drop production in Sec.~\ref{sec:mc}

\subsection{Statistical analysis of raft decay}\label{sec:stat_decay}
\begin{figure}[htbp]
    \centering
    \includegraphics[width=\linewidth]{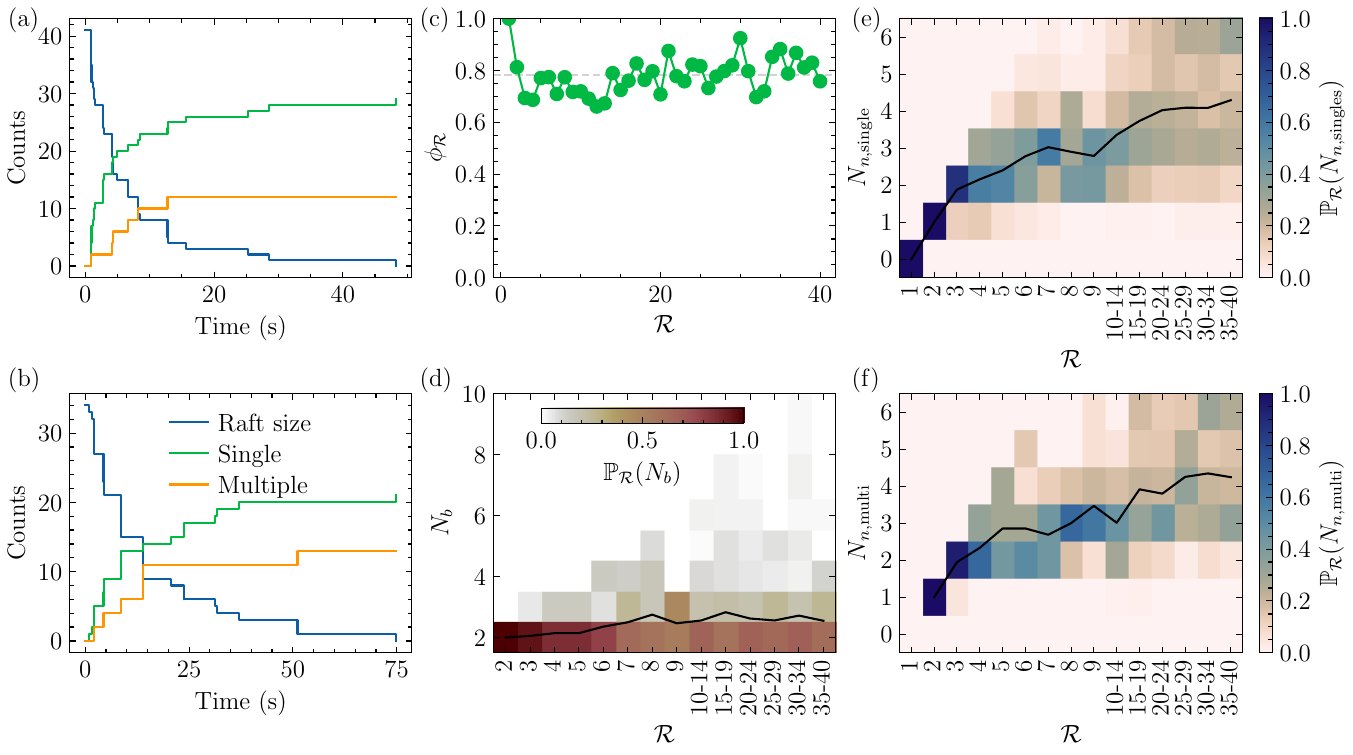}
    \caption{Examples and statistical properties of raft decay. (a) and (b) show time evolutions of the number of bubbles in two rafts with 13.6 and 27.2 \textmu mol/L of SDS, respectively. Green and orange curves show events attributed to single and multiple bursting events, respectively. (c) Evolution of the proportion of single bursting events $\phi_\mathcal{R}$ with raft size. (d) Heat map of the probability density of the number of bursting bubbles in a multiple bursting event, as a function of raft size. The black line shows the mean value. (e) and (f) show the probability density of the number of neighbors of the bursting bubble, as a function of raft size, with the black line representing the mean number of neighbors. (e) corresponds to single bursting events and (f) to multiple events. In the latter case, the number of neighbors reported is the one corresponding to the first bursting bubble.}
    \label{fig:raft_stats}
\end{figure}

In this section, we analyze the prevalence of each type of bursting event (single or multiple) and measure statistical properties of raft decay. This is done using a setup that includes only the top view (described in Sec.~\ref{sec:setup_stat}), allowing us to track raft decay for a large number of events (86 individual rafts and 2100 bursting bubbles). Two examples of raft decay are shown in Fig.~\ref{fig:raft_stats} (a) and (b). The number of bubble decreases over a time which is typically of the order of 1 minute, with a large number of bursting events just after generation. Most of the events are single bursting events, but one multiple bursting event can contain up to 10 bubbles at the same time. The resulting data is automatically categorized into single and multiple bursting events using the criteria described above and summarized in Fig.~\ref{fig:schematic_multi}.

The frequency of single bursting events $\phi_\mathcal{R}$ is reported in Fig.~\ref{fig:raft_stats} (c). Note that while a unique multiple bursting event includes several bubbles, it is still counted as only one event. In contrast, two simultaneous but physically separated single bursting events are counted as two different events. The frequency is strikingly constant across all raft sizes with single burstings representing roughly 80\% of events. As the raft size approaches one, the frequency of single bursting events rises to one.

We can now further analyze each type of event: for the multiple bursting case, we compute the probability density of the number of bursting bubbles $N_b$ for a given raft size: $\mathbb{P}_\mathcal{R}(N_b)$. This probability is represented as a heat map for all raft sizes in Fig.~\ref{fig:raft_stats} (d). The number of simultaneously bursting bubbles is typically between 2 and 4, with extremes potentially reaching up to 10 bubbles. The typical number of bursting bubbles increases with raft size, up to $\mathcal{R} \approx 9$, and is relatively constant around 2.5 afterwards.  

We finally report the probability distribution of the number of neighbors $\mathbb{P}_\mathcal{R}(N_n)$ in each case in Fig.~\ref{fig:raft_stats} (e) and (f). In the single bursting event case (panel (e)), the number of neighbors $N_{n,\,\mathrm{single}}$ increases gradually, from 0 to around 3 neighbors as $\mathcal{R}$ increases from 1 to 4. For $4 < \mathcal{R} < 10$, low neighbor counts become rare as they would require chain-like configurations, and very high counts ($N_{n,\,\mathrm{single}} \geqslant 5$) are also infrequent. From $\mathcal{R} = 10$, the distribution of neighbor counts shifts, and all values from 3 to 6 become likely. 

In the case of multiple bursting events (panel (f)), we only report the neighbor count of the first bursting bubble $N_{n,\,\mathrm{multi}}$ as we have shown in the previous section that the jetting dynamics is independent of the number of neighbors for the subsequent bursting bubbles. The distribution of neighbor counts $\mathbb{P}_\mathcal{R}(N_{n,\,\mathrm{multi}})$ is slightly different compared to the previous case: a raft size of 1 or 0 neighbors are now impossible configurations, but the average neighbor counts are very similar. 

\subsection{Estimation of raft drop production}\label{sec:mc}

\begin{figure}[htbp]
    \centering
    \includegraphics[width=0.8\linewidth]{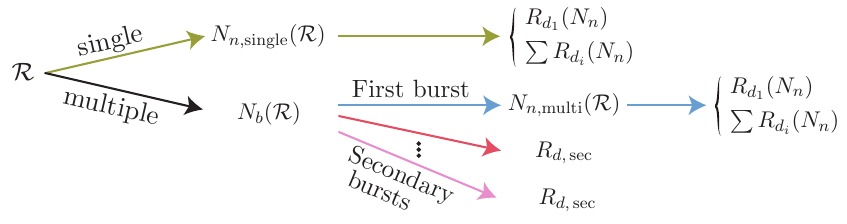}
    \caption{Schematic of the raft distribution estimation. For all single bursting events, the number of neighbors is selected, and the drop size distribution is then selected accordingly. Two options are possible: counting only the first drop or all drops generated per event.
    In the case a multiple bursting event, the number of bursting bubbles is selected first. For the first bursting bubble, the drop size selection is identical to the single bursting case. For secondary bubbles, the drop sizes are not correlated with the number of neighbors.}
    \label{fig:mc_schematic}
\end{figure}

Using the raft decay and bursting statistics presented above, we now detail a method to estimate the drop size distribution associated with a raft of monodisperse bubbles. This estimator takes as input the raft decay statistics in panels (c) to (f) of Fig.~\ref{fig:raft_stats}: the fraction of single bursting events $\phi_\mathcal{R}$, the probability distribution of number of bursting bubbles in the case of a multiple bursting event $\mathbb{P}_\mathcal{R}(N_b)$, and the probability distributions of number of neighbors in the single and multiple cases $\mathbb{P}_\mathcal{R}(N_{n,\,\mathrm{single}})$ and $\mathbb{P}_\mathcal{R}(N_{n,\,\mathrm{multi}})$. In addition, the estimator requires the distribution of drop sizes for a given number of neighbors in the single bursting event case (Fig.~\ref{fig:first_drop_rad}) and in the multiple bursting case (Fig.~\ref{fig:data_multi}). We then obtain as outputs for each raft size, the drop size distribution relative to the case with no neighbors $\mathbb{P}_\mathcal{R}(\hat{R_d})$, with $\hat{R_d} = R_{d} / R_{d}(N_n=0)$. This drop size distribution corresponds to the number of drops of a given size produced per bursting event. 

The working principle of the estimator is represented schematically in Fig.~\ref{fig:mc_schematic}. Given a raft size $\mathcal{R}$, the type of bursting event is can either be single or multiple. The probability of a single bursting event is $\phi_R$, following Fig.~\ref{fig:raft_stats} (c). Two cases are then possible:
\begin{itemize}
    \item In the case of a single bursting event, the number of neighbors of the single bursting bubble $N_{n,\,\mathrm{single}}(\mathcal{R})$ follows the distribution in Fig.~\ref{fig:raft_stats} (e). Given the number of neighbors, the drop size is then finally follows the distribution $\mathbb{P}_{N_n}(\hat{R_d})$. 
    \item In the case of multiple bursting events, the number of bursting bubbles $N_b$ follows the distribution in Fig.~\ref{fig:raft_stats} (d). The first bursting bubble is then treated similarly to the single bursting event case: $N_{n,\,\mathrm{multi}}(\mathcal{R})$ is drawn, and the drop size follows from $\mathbb{P}_{N_n}(\hat{R_d})$.
    For the secondary bursting bubbles, no neighbor count is drawn, and the drop size directly follows the distribution of secondary bursting events.
\end{itemize}
The algorithm described above can be implemented as a Monte-Carlo simulation of raft decay, but it is simple enough that an analytical derivation is actually possible. 

Mathematically, the drop size distribution $\mathbb{P}_\mathcal{R}(\hat{R_d})$ for a raft of size $\mathcal{R}$ is derived as follows. The distribution decomposed into single and multiple bursting events using the fraction of single bursting events $\phi_{\mathcal{R}}$ through:
\begin{equation}
    \mathbb{P}_\mathcal{R}(\hat{R_d}) = \phi_{\mathcal{R}} \mathbb{P}_{\mathcal{R}, \rm single}(\hat{R_d}) +  (1 -  \phi_{\mathcal{R}}) \mathbb{P}_{\mathcal{R}, \rm multi}(\hat{R_d}),
\end{equation}
with $\mathbb{P}_{\mathcal{R}, \rm single}(\hat{R_d})$ the drop size distribution from single bursting events only, and $\mathbb{P}_{\mathcal{R}, \rm multi}(\hat{R_d})$ the distribution from multiple bursting events only. 

The distribution in the case of single bursting events only is described in Eq.~\eqref{eq:p_single} and is the combination of the distribution of number of neighbors for the given raft size $\mathbb{P}_{\mathcal{R}, \rm single}(N_n)$ (plotted in Fig.~\ref{fig:raft_stats} (e)) and the drop size distribution for a bursting bubble with $N_n$ neighbors $\mathbb{P}_{N_n}(\hat{R_d})$. 

The case of multiple bursting events is detailed in Eq.~\eqref{eq:p_multi}: we consider all possible number of bursting bubbles along with their distribution (Fig.~\ref{fig:raft_stats} (d)), then in addition to considering the first bubble in the same way as in the previous case, we also consider the $N_b - 1$ other bursting bubbles, each with a drop size distribution $\mathcal{P}_{\rm sec}(\hat{R_d})$. Note that with the way we have normalized the distributions, they represent a number of drops per bursting event. 
\begin{subequations}
    \begin{align}
        \mathbb{P}_{\mathcal{R}, \rm single}(\hat{R_d}) & = \sum_{N_n=0}^{6} \mathbb{P}_{\mathcal{R}, \rm single}(N_n) \mathbb{P}_{N_n}(\hat{R_d}), \label{eq:p_single}\\ 
        \mathbb{P}_{\mathcal{R}, \rm multi}(\hat{R_d}) & =  \sum_{N_b} \mathbb{P}_\mathcal{R}(N_b) \left(\sum_{N_n=0}^{6} \mathbb{P}_{\mathcal{R}, \rm multi}(N_n) \mathbb{P}_{N_n}(\hat{R_d}) + (N_b - 1) \mathcal{P}_{\rm sec}(\hat{R_d}) \right).\label{eq:p_multi}
    \end{align}
\end{subequations}
Finally, we specify the drop size distributions associated with a bubble with $N_n$ neighbors $\mathbb{P}_{N_n}(\hat{R_d})$. We denote with $\mathcal{P}$ the drop size distributions used as inputs of the estimator. We can either consider only the first ejected drop, in which case the distribution is simply $\mathcal{P}_{N_n}^{\rm first}(\hat{R_d})$, a distribution fitted onto the data from Fig.~\ref{fig:first_drop_rad} (Eq.~\eqref{eq:pfirst}). We can also consider all drops ejected: rigorously, this would require knowing the size distribution of the ith drop, for each number of neighbors (Eq.~\eqref{eq:pall}), along with the distribution of number of drops $\mathbb{P}_{N_n}(N_d)$. Since our experiments do not give us access to this type of precision, we make the hypothesis that the distribution for all drops from a given bursting event is the same. As a consequence, we only need the combined distribution for all drops $\mathcal{P}_{N_n}^{\rm all}(\hat{R_d})$ and the average number of drops $\left< N_d \right>_{N_n}$.
\begin{subequations}
    \begin{align}
    \mathbb{P}_{N_n}^{\rm first}(\hat{R_d}) & = \mathcal{P}_{N_n}^{\rm first}(\hat{R_d}), \label{eq:pfirst}\\
    \mathbb{P}_{N_n}^{\rm all}(\hat{R_d}) & = \sum_{N_d} \mathbb{P}_{N_n}(N_d) \sum_{i=1}^{N_d} \mathbb{P}_{N_n}(R_{d, i}) \approx \left< N_d \right>_{N_n} \mathcal{P}_{N_n}^{\rm all}(\hat{R_d}). \label{eq:pall}
    \end{align}
\end{subequations}

\begin{figure}[htbp]
    \centering
    \includegraphics[width=0.87\linewidth]{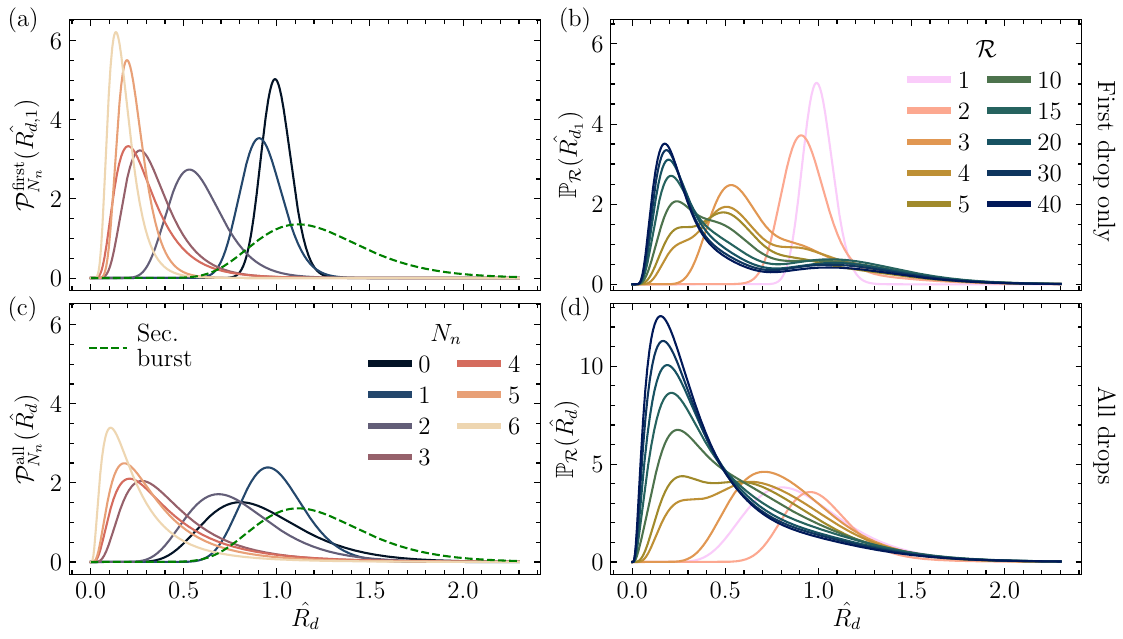}
    \caption{Inputs and outputs of the drop size distribution estimator. (a) Fitted distributions of drop size obtained from the data on individual bubbles. All distributions are log-normals, and the dashed green line corresponds to secondary bursting bubbles in multiple bursting events. (b) Drop size distributions obtained from raft of a given size $\mathcal{R}$. The curves are obtained from a single step of the Monte-Carlo simulation. Panels (c) and (d) are identical to panels (a) and (b), respectively, with the difference that all drops are included in the distributions instead of the first one only.}
    \label{fig:mc_pdfs}
\end{figure}

We fit the experimental data of drop sizes with distributions to improve the regularity of the output. We use log-normal distributions in all cases, that we fit on the data in Fig.~\ref{fig:first_drop_rad}, combining all data per neighbor count to fit $\mathcal{P}_{N_n}^{\rm first}(\hat{R_d})$. For the secondary bursts in the multiple bursting events case, we group all the data in Fig.~\ref{fig:data_multi} (a) with burst order larger than one and also use a log-normal distribution to fit $\mathcal{P}_{\rm sec}(\hat{R_d})$. Note that use of log-normal distributions is arbitrary and that other distributions on positive values such as the Gamma distribution have also been used to describe drop size distributions \citep{villermaux_fragmentation_2020,deike_mechanistic_2022}. The fitted distributions are shown in Fig.~\ref{fig:mc_pdfs} (a) for varying neighbor counts $N_n$ (colors). All distributions are relatively narrow, except for the secondary bursting events (dashed green). 

The collective effects on jet drop production are still very visible when considering the drop size distribution of a raft $\mathbb{P}_\mathcal{R}(\hat{R_{d_1}})$ (see Fig.~\ref{fig:mc_pdfs}, (b)). For $\mathcal{R} = 1$, the resulting distribution is identical to the distribution used as input to the model for $N_n=0$. Note that this distribution is also the one that would be obtained when considering a raft of non-interacting monodisperse bubbles and therefore serves as a reference case. As $\mathcal{R}$ increases, the distribution both shifts to smaller sizes and gets wider. Smaller sizes appear in the distribution due to the probability of single bursting event with large neighbor counts increasing, and therefore contributing to smaller drop sizes. The widening of the drop size distribution is in part due to the widening of the distribution of possible neighbor counts, and to the large width of the distribution for secondary bursting events, that get increasingly likely as $\mathcal{R}$ increases. Finally, from $\mathcal{R} \approx 30$ the distribution approaches a limit and only weakly depends on raft size. The resulting drop size distribution has two distinct peaks: a small one roughly at the size of isolated drops, due to secondary bursting events and one at the size corresponding to bursting with six neighbors, roughly at $\hat{R_{d_1}} = 0.2$. Note that the integral of the distributions correspond to the number of bubbles bursting on average per event, which approaches 1.3 for large raft sizes.

The calculations can be adapted to consider all drops produced instead of the first one only. This is accomplished by fitting the input distributions of drop size for a given neighbor count $\mathcal{P}_{N_n}^{\rm all}(\hat{R_d})$ on all drops measured with the high speed imaging instead of the first one only like in the previous case (see Fig.~\ref{fig:mc_pdfs} (c)). The data for the size of all drops as a function of the number of neighbors can be found in App.~\ref{app:all_drops}. The fitted distributions are generally more broad than when considering the first drop only, and their average shift slightly. In particular, the size distribution for all drops in the isolated case (black) has a mean around 0.75, meaning that it is smaller than the mean size of the first drop. In contrast, the case with 1 neighbor has a larger mean when considering all drops than when considering only the first one.
The resulting drop size distribution for a given raft size is shown in Fig.~\ref{fig:mc_pdfs} (d). These distributions are normalized so that their integrals corresponds to the average number of drops ejected for a given bursting event. As a result, the number of drops per bubble can be seen to increase as raft size increases because large neighbor counts generate many small drops. The size reduction effect therefore appears more strongly than when considering only the first drop. For large raft sizes, the peak around 1 disappears as secondary bursting does not produce many drops. Instead, the distribution exhibits a long tail towards large sizes. The number of drops per bursting event increases from 2.5 for isolated bubbles to almost 6 for large raft sizes.

Regardless of the specifics of the simulation, robust trends can be observed from this statistical approach derived from the experimental properties of bubbles bursting in rafts: when bubbles burst in rafts, collective effects on the bursting process are enough to significantly broaden the drop size distribution and reduce the typical drop size compared to what would be observed for bubbles bursting in isolation. Integrating the distribution in the case with all drops for a raft size of 40, we find that the average drop size is 40\% of what it would be for bubbles in isolation. We have further strengthened this result by simulating the trends obtained when considering the full raft decay, or when changing the statistical properties of raft decay, by allowing longer range interactions between bursting bubbles, with details of these sensitivity tests given in App.~\ref{app:other_mc}. .

\section{Discussion}\label{sec:conclusion}

\begin{figure}[htbp]
    \centering
    \includegraphics[width=0.8\linewidth]{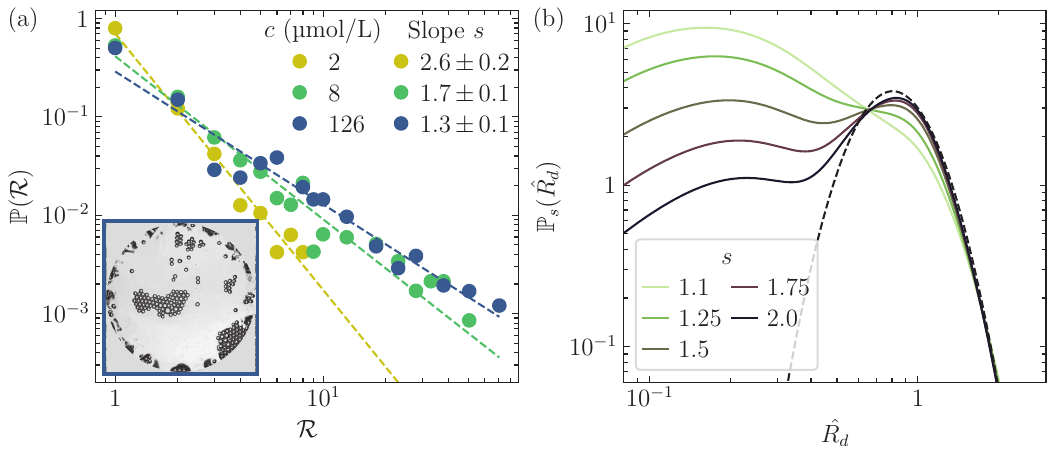}
    \caption{(a) Raft size distribution found in the data from \citet{neel_role_2022}, for various SDS concentrations (colors). The inset shows an example image with rafts from that experiment. Dashed lines are fits of the data following a power law distribution with exponent $s$. (b) Drop size distribution resulting from a range of raft sizes, distributed according to a power law of exponent $s$. The dashed line corresponds to a case with isolated bubbles.}
    \label{fig:bapt_noe}
\end{figure}

In this study we have analyzed the effect of neighboring bubbles on the production of droplets through the jet drops mechanism. For surface bubbles in the high Laplace number regime, we have shown that neighbors strongly alter jet formation, forming smaller, faster and more numerous drops as the neighbor count is increased up to six for hexagonal packing. We attribute this size reduction to an increase in dissipation during cavity collapse, resulting in the selection of a thinner and faster jets.

We have also analyzed cases where multiple bubbles burst in quick succession, each triggered by previous bursting events. In this case, we found that while the first bursting bubble is unaffected by secondary bursting events, the dependence of drop size with neighbor geometry is lost for those secondary bursting events. We constructed a framework to estimate raft drop production using statistical properties of raft decay, and we showed that collective effects in bubble rafts significantly reduce the typical drop size and increase the width of the drop size distribution compared to a case with no interactions. 

Still, the case with a single raft size is an idealization of real situations where bubbles aggregate in rafts (at the surface of sparkling drinks or the ocean). In practice, the drop size distribution that would be measured in those situations would result from many different raft sizes simultaneously. We can estimate this combined drop size distribution $\mathbb{P}(\hat{R_d})$ as the result of the sum of the distribution for a given raft size $\mathbb{P}_\mathcal{R}(\hat{R_d})$ that we have previously computed, multiplied by the probability of that raft size $\mathbb{P}(\mathcal{R})$:
\begin{equation}\label{eq:convol_raft}
    \mathbb{P}(\hat{R_d}) = \sum_{\mathcal{R}} \mathbb{P}(\mathcal{R}) \: \mathbb{P}_\mathcal{R}(\hat{R_d}) .
\end{equation}
To probe how our work would apply in a more realistic case, we reanalyzed surface bubble data from \citet{neel_role_2022}. In this study, the authors continuously generate many bubbles in a large tank and measure the surface bubble size distribution as well as the drop size distribution, while varying the surfactant concentration. As the surfactant concentration increases, the authors observe the bubble forming nearly monodisperse rafts, and the drop size distribution increasing in number and becoming much wider than in the low contamination case.

We can qualitatively reproduce the same result by using the distributions obtained for a single raft size with the Monte-Carlo simulation and measuring the raft size distribution from the data of \citet{neel_role_2022}. An example image from this study is reproduced in the inset of Fig.~\ref{fig:bapt_noe}. The raft size distribution is found to be well described by a power law distribution for all surfactant concentrations (Fig.~\ref{fig:bapt_noe} (a)): $\mathbb{P}(\mathcal{R}) \propto \mathcal{R}^{-s}$, with an exponent $s$ between $1$ and $2.7$ that decreases with surfactant concentration (i.e. large rafts get more likely). Using this experimentally measured raft size distribution, we are able to assess, using Eq.~\eqref{eq:convol_raft}, the effect of neighbors on drop size distributions in the case with a broad distribution of raft sizes. The result of this operation is shown in Fig.~\ref{fig:bapt_noe} (b) with various exponents $s$ between 1 and 2. We finally see in this general case, that simply by including collective effects on bubble collapse, the drop size distribution becomes much wider, with a peak value that shifts to a smaller size. Increasing the exponent $s$ decreases the likelihood of large rafts and therefore reduces collective effects, approaching the size distribution corresponding to isolated bubbles (dashed line). These observations point towards a significant fraction of the trends seen in \citet{neel_role_2022} being due to collective effects. In particular, the location of the peak of the distribution in this study is reduced by a factor around 5 compared to where it would be if bubbles burst in isolation, which is very close to what we can observe from our estimated distributions.

The discovery of this new size reduction effect through neighbor-induced dissipation opens several new and important questions to better understand how raft formation affects drop production in general. The most pressing of these questions is to put bounds on the size difference between neighboring bubbles for the size reduction to occur. In this study, we have used narrow-banded distributions of bubble sizes (less than 10\% variations in $R_b$ in a raft), but in many applications (such as in the ocean) bubbles sizes at the surface can span several orders of magnitude \citep{deike_mechanistic_2022}. Intuitively it seems clear that large size discrepancies will limit the effects described above, but investigating the magnitude of collective effects in broad-banded rafts appears crucial to better describe drop production. 

In addition to this challenge, studying how the size reduction is altered in different regions of the parameter space for jet drops appears to be an interesting avenue. In particular two cases would allow to understand how ubiquitous collective effects are: the first one is the lower Laplace number regime ($\mathrm{La} \sim 10^3$) where dissipation already produces very thin and fast jets. This regime would help define a lower bound for the bubble size where collective effects can occur. The second interesting perspective is at large surfactant concentrations: the ones used in this study are very small on purpose, but larger concentration can significantly alter jetting, producing larger or smaller drops depending on the Laplace number \citep{pierre_influence_2022,vega_influence_2024,eshima_size_2025,rodriguez-aparicio_critical_2025}. Studying collective effects with contaminated interfaces would therefore allow using the conclusions from this study in an even broader range of industrial or geophysical applications.

The code used to generate Fig.~\ref{fig:mc_pdfs} is publicly available at \url{https://github.com/DeikeLab/drop_distribution_neighbors}
\begin{acknowledgments}
This work was supported by NSF grant 2242512 and 2318816 to L.D. N.D acknowledges support from Institut Philippe Meyer. 
\end{acknowledgments} 

\bibliography{bubbles}

\appendix



\section{Evolution of the size of all drops}\label{app:all_drops}

\begin{figure}[htbp]
    \centering
    \includegraphics[width=0.9\linewidth]{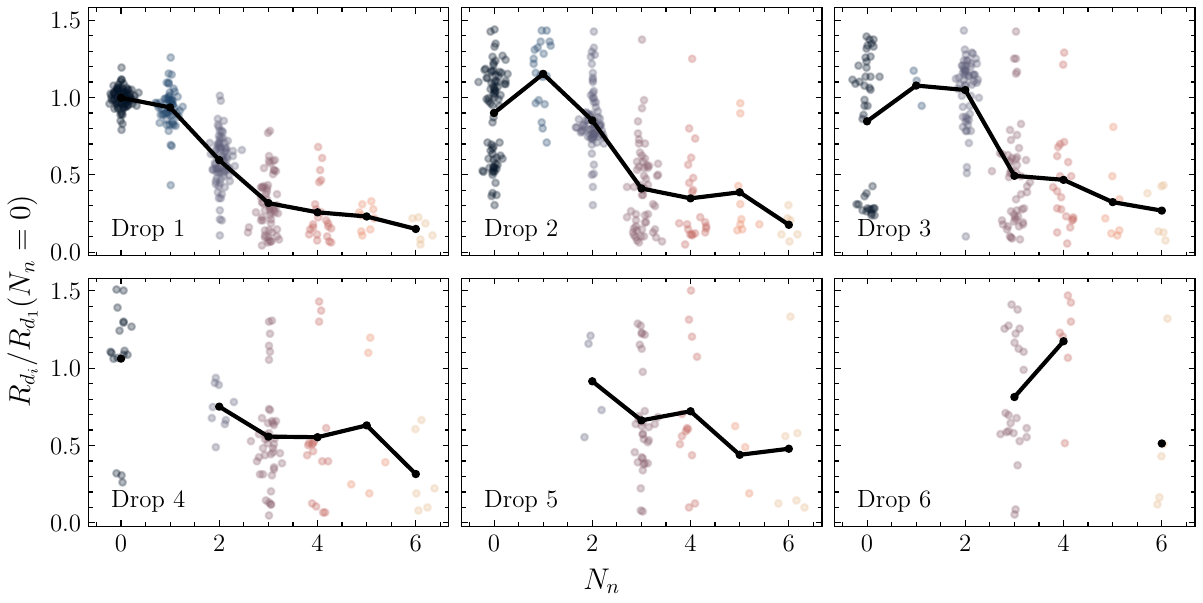}    
    \caption{Evolution of dimensionless drop size with number of neighbors. Each panel corresponds to a drop number. Points are individual drops and the black line is the average. }
    \label{fig:all_drops}
\end{figure}

This section is dedicated to the analysis of the drops beyond the first one. This data is used to fit the distributions shown in Fig.~\ref{fig:mc_pdfs} (c). The drop size is shown in Fig.~\ref{fig:all_drops} as a function of neighbor count, and each panel corresponds to a drop number (so that the drop 1 panel shows the same data as Fig.~\ref{fig:first_drop_rad}). This figure combines data by normalizing the radius of each drop $R_{d_i}$ by the mean radius of the first drop for each surfactant concentration or bubble size. 

The very clear trend of smaller drops sizes observed for the first drop in the main text is still visible for subsequent drops where as $N_n$ is increased, $R_{d_i}$ generally decreases regardless of the drop number $i$. For drops number 2 and 3 the drop size actually increases first for 1 or 2 neighbors to values above $R_{d_1}(N_n=0)$, before decreasing for larger neighbor counts. As the drop number increases, there is no more data for low neighbor counts as they don't typically produce enough drops. 

We can also note in this view that for the isolated case $N_n = 0$, the second and third drops ejected have two distinct peaks at different sizes: one close to $\hat{R_d}$ = 1 and the other to 0.5 and 0.25, respectively. This forking behavior: having a well define first drop size and subsequent drops alternating between small and large size, was previously observed numerically by \citet{berny_size_2022}.

\section{Variations of the drop size distribution estimator}\label{app:other_mc}

\begin{figure}[htbp]
    \centering
    \includegraphics[width=\linewidth]{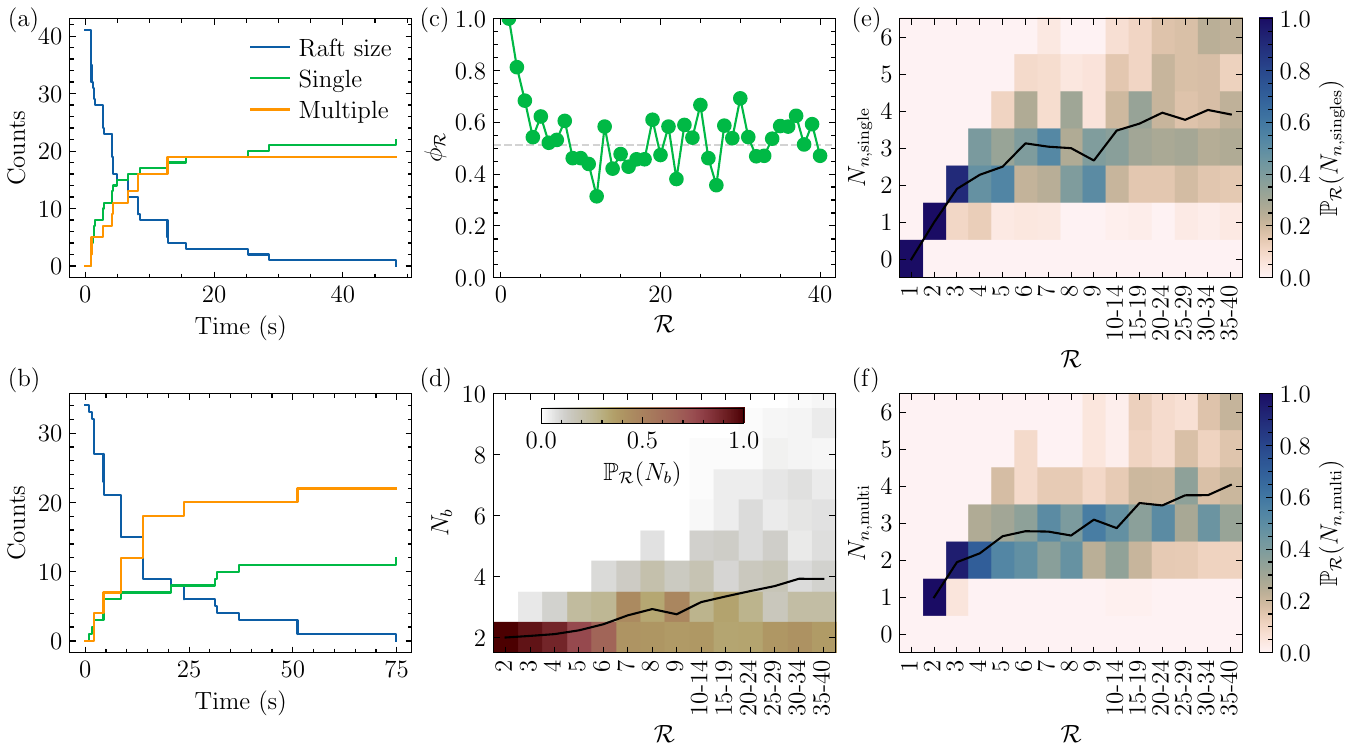}
    \caption{Effect of long range interactions on the statistical properties of rafts. (a) and (b) show time evolutions of the number of bubbles in two rafts with 13.6 and 27.2 \textmu mol/L of SDS, respectively. Green and orange curves show events attributed to single and multiple bursting events, respectively. (c) Evolution of the proportion of single bursting events with raft size $\phi_\mathcal{R}$. (d) Heat map of the probability density of the number of bursting bubbles in a multiple bursting event, as a function of raft size. The black line shows the mean value. (e) and (f) show the probability density of the number of neighbors of the bursting bubble, as a function of raft size, with the black line representing the mean number of neighbors. (e) corresponds to single bursting events and (f) to multiple events. In the latter case, the number of neighbors reported is the one corresponding to the first bursting bubble.}
    \label{fig:raft_stats_4p5}
\end{figure}

\begin{figure}[htbp]
    \centering
    \includegraphics[width=0.9\linewidth]{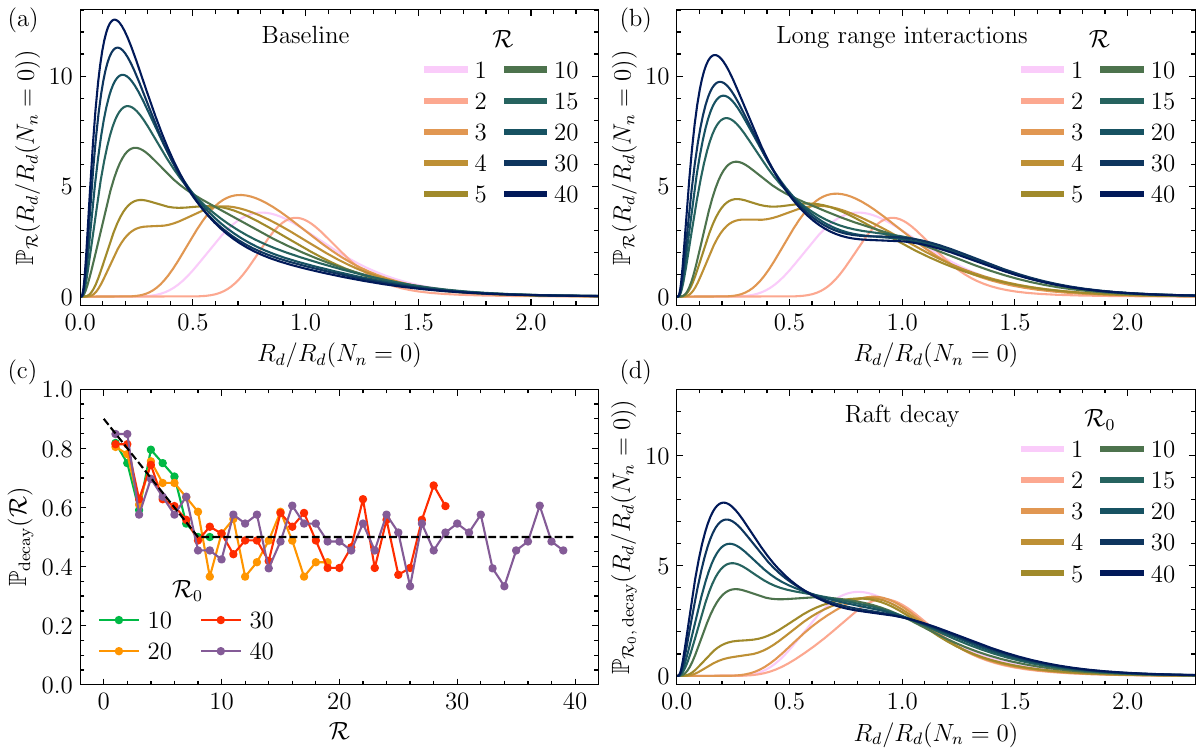}
    \caption{Effect of parameters on simulated drop size distributions. (a) Baseline case with all drops considered (identical to Fig.~\ref{fig:mc_pdfs} (d)). (b) Identical to (a) but assuming long range interactions between bubbles, i.e. not only neighbors but also bubbles adjacent to neighbors are considered secondary bursting events. (c) Drop size distribution resulting from the complete decay of a raft of initial size $\mathcal{R}_0$. }
    \label{fig:other_mc}
\end{figure}

We used the Monte-Carlo simulation described in Sec.~\ref{sec:mc} to assess the effect of our assumptions or the choice of parameters in the simulations. In Fig.~\ref{fig:other_mc} (a) we reproduced the case where all drops are considered from the main text (i.e. panel (d) from Fig.~\ref{fig:mc_pdfs}) to compared it to the two main variations of the simulations. 

The first case is allowing for longer range interactions between bubbles. In the main text, we assume that two bursting events belong to the same sequence only if they are directly adjacent. We now extend this definition to also include bubbles only one neighbor away. In practice, this means changing the maximum distance parameter $\delta_{\rm max}$ defined in Sec. \ref{sec:setup} from 2.3 to 4.5: in a hexagonal lattice, bubbles one neighbor away from the bursting bubble are at a dimensionless distance $2\sqrt3 \approx 3.5$ or $4$, while the next closest bubble is at a distance 5. 

This change impacts the parameters of the estimator that are measured from the raft decay data (panels (c) to (f) of Fig.~\ref{fig:raft_stats}). As a result, we reproduce the exact equivalent of Fig.~\ref{fig:raft_stats} in Fig.~\ref{fig:raft_stats_4p5}, only changing the value of $\delta_{\rm max}$. The most direct impact is on the probability of a single bursting event (panel (c)) that decreases from roughly 80\% of all events to 50\%, independently of raft size. Conversely, the number of bursting bubble per multiple bursting event $N_b$ increases to match the total number of bursting bubbles. There is only a minor impact on the number of neighbors in either case ($N_{n,\mathrm{single}}$ and $N_{n,\mathrm{multi}}$).
The result of the drop size distribution estimator calculation with these new parameters is shown in Fig.~\ref{fig:other_mc} (b), considering all ejected drops. The trends of shifting towards smaller sizes and widening of the distribution observed on the short-range interaction case are still present. However, due to the larger relative proportion of multiple bursting events, the peak at small size for large rafts is reduced and is associated with an increase of the peak corresponding with secondary bursting events around $R_d/R_d(N_n=0)=1$. 

A second extension of the estimator presented in Sec.~\ref{sec:mc} is to predict the drop size distribution resulting from the decay of a whole raft until every bubble in it has burst instead of focusing on a raft of given size. Within our purely statistical framework with no sense of time evolution, we estimate the decay distribution from a raft of size $\mathcal{R}_0$ by summing the distributions associated with raft sizes below $\mathcal{R}_0$ weighted accordingly, similarly to the case for many different raft sizes (Eq.~\eqref{eq:convol_raft}):
\begin{equation}\label{eq:decay}
    \mathbb{P}_{\mathcal{R}_0, \rm decay}(\hat{R_d}) = \frac{1}{\mathcal{R}_0} \sum_{\mathcal{R} \leqslant \mathcal{R}_0} \mathbb{P}_{\mathcal{R}_0, \rm decay}(\mathcal{R}) \mathbb{P}_\mathcal{R}(\hat{R_d}).
\end{equation}
The probability $\mathbb{P}_{\mathcal{R}_0, \rm decay}(\mathcal{R})$ corresponds to the probability of visiting the raft size $\mathcal{R}$ when starting from a raft size $\mathcal{R}_0$. Not all raft sizes are visited because of multiple bursting events, and we can measure this probability from our experiments of raft decay (Fig.~\ref{fig:raft_stats} (a) and (b)). This probability \emph{a priori} depends on the initial raft size $\mathcal{R}_0$, and we plot its evolution in Fig.~\ref{fig:other_mc} (c) for several initial raft sizes (and for short range interactions). We can actually see that it is independent of $\mathcal{R}_0$ and approximately equal to 0.5 for $\mathcal{R} > 10$. The probability is decreasing from $\mathcal{R} = 0$ to 1 as multiple bursting events become increasingly likely. We therefore compute $\mathbb{P}_{\mathcal{R}_0, \rm decay}(\hat{R_d})$ using Eq.~\eqref{eq:decay} and with the fit $\mathbb{P}_{\rm decay}(\mathcal{R}) = {\rm max}(0.9 - \mathcal{R} / 20, 0.5)$ (black dashed line).

The resulting drop size distribution in the case taking into account all drops and with short range interactions, is shown in Fig.~\ref{fig:other_mc} (d). Note that the distribution $\mathbb{P}_{\mathcal{R}_0, \mathrm{decay}}(\hat{R_d})$ represents the number of drops ejected per bubble initially present in the raft, to be able to compare different raft sizes. Once again, the trends visible in the main text are unchanged by this different approach. Considering the decay of a raft instead of a single step reinforces the peak around 1, as all rafts end up having only few bubbles and therefore sample low neighbor counts more often.

\end{document}